\documentclass[a4paper]{article}

\usepackage{INTERSPEECH2020}
\usepackage{graphicx,xcolor}
\usepackage{slashbox}
\usepackage{multirow}

\title{Joint Speaker Counting, Speech Recognition, and Speaker Identification for Overlapped Speech of Any Number of Speakers}

\name{Naoyuki Kanda, Yashesh Gaur, Xiaofei Wang, Zhong Meng, \\Zhuo Chen, Tianyan Zhou, Takuya Yoshioka}
\address{
  Microsoft Corp.} 
\email{\{Naoyuki.Kanda,Yashesh.Gaur,Xiaofei.Wang,Zhong.Meng,zhuc,tizhou,tayoshio\}@microsoft.com}

\begin{document}

\maketitle
\begin{abstract}
We propose
an end-to-end speaker-attributed automatic speech recognition model that
unifies
speaker counting, speech recognition, and speaker identification on monaural overlapped speech.
Our model is built on serialized output training (SOT) with attention-based encoder-decoder, a recently proposed method 
for recognizing overlapped speech comprising an arbitrary number of speakers. 
We extend SOT by introducing
a speaker inventory as an auxiliary input 
to produce
speaker labels as well as multi-speaker transcriptions.
All model parameters 
are optimized by 
speaker-attributed maximum mutual information criterion,
which represents a joint probability for 
overlapped speech recognition
and speaker identification.
Experiments on LibriSpeech corpus show that 
our proposed method achieves significantly better 
speaker-attributed word error rate than
the baseline
that separately performs 
overlapped speech recognition and speaker identification. 
\end{abstract}

\section{Introduction}

Speaker-attributed automatic speech recognition (SA-ASR)
from overlapped speech
has been an active research area
towards meeting transcription \cite{fiscus2007rich,janin2003icsi,carletta2005ami}.
It requires to count the number of speakers,
transcribe utterances that are sometimes overlapped, and also diarize or identify the speaker of 
each utterance.
While significant progress has been made especially for multi-microphone settings (e.g., \cite{yoshioka2019advances}), 
SA-ASR remains very challenging when we can only
access monaural audio. 

A significant amount of research has been 
conducted 
to achieve the goal of SA-ASR. 
One approach is applying speech separation (e.g., \cite{hershey2016deep,chen2017deep,yu2017permutation})
before ASR and speaker diarization/identification. 
However, a speech separation module is often designed with
a signal-level criterion, 
which is not necessarily optimal for 
succeeding 
modules. 
To overcome this suboptimality, 
researchers have investigated approaches for jointly 
modeling multiple modules. 
For example, there are a number of studies  
concerning joint modeling of 
speech separation and ASR
(e.g., \cite{yu2017recognizing,seki2018purely,chang2019end,chang2019mimo,kanda2019acoustic,kanda2019auxiliary}).
Several methods were also proposed for integrating speaker identification and speech separation~\cite{wang2019speech,von2019all,kinoshita2020tackling}. However, little research has yet been done to address
SA-ASR by combining \emph{all} these modules. 

Only a limited number of studies
 have tackled
 the joint modeling of  
multi-speaker ASR and speaker diarization/identification.
\cite{el2019joint} proposed to generate transcriptions of 
different speakers interleaved by speaker role tags 
to recognize two-speaker conversations based on
a recurrent neural network transducer (RNN-T). 
Although promising results were shown 
for two-speaker conversation 
data,
the method cannot deal with speech overlaps
due to the monotonicity constraint of RNN-T. 
Furthermore,
their method is difficult to be extended to  
an arbitrary number of speakers
because the speaker role tag needs to be uniquely defined for each speaker (e.g., a doctor and a patient). 
\cite{kanda2019simultaneous} proposed a
joint decoding framework for
overlapped speech recognition and speaker diarization,
in which 
speaker embedding estimation and target-speaker ASR were applied 
alternately.
Although this method is extendable to many speakers in theory,
it assumes that 
the speaker counting is conducted during the speaker embedding estimation
process, which is challenging in practice.

In this paper,
we propose 
an end-to-end
SA-ASR 
model
that 
unifies
speaker  counting,
overlapped speech  recognition,  and  speaker  identification.
 Our model is built on serialized output training (SOT) \cite{kanda2020sot} with attention-based encoder-decoder (AED) 
 \cite{bahdanau2014neural,chorowski2014end,chorowski2015attention,chan2016listen}, 
 which was recently proposed for recognizing overlapped speech
 consisting of an arbitrary number of speakers.
We extend the SOT model  by introducing 
a speaker inventory as an auxiliary input 
to produce 
speaker labels as well as multi-speaker transcriptions.
All model parameters 
are optimized by maximizing a joint probability for 
overlapped speech recognition and speaker identification.
Our model can recognize
overlapped speech of {\it any number of speakers} 
while identifying the speaker of
each utterance among {\it any number of speaker profiles}.
We show that the proposed model achieves 
 significantly  better  speaker-attributed  word error rate (SA-WER)
over the model consisting of separate modules.

\section{Overlapped Speech Recognition with Serialized Output Training}
\subsection{ASR based on Attention-based Encoder Decoder}

Given input $X=\{x_1,...,x_T\}$, an AED model produces 
a posterior probability of output sequence $Y=\{y_1, ..., y_n, ..., y_N\}$ as follows. 
Firstly, an encoder converts the input sequence $X$ 
into a sequence, $H^{enc}$, of embeddings, i.e.,  
 \begin{align}
 H^{enc} &=\{h^{enc}_1,...,h^{enc}_T\}={\rm AsrEncoder}(X).  \label{eq:enc} 
\end{align}
Secondly, at each decoder step $n$, the attention module 
outputs attention weight $\alpha_n =\{\alpha_{n,1},...,\alpha_{n,T}\}$
as
\begin{align}
 \alpha_n &= {\rm Attention}(u_n, \alpha_{n-1}, H^{enc}), \label{eq:att} \\
 u_n &={\rm DecoderRNN}(y_{n-1}, c_{n-1}, u_{n-1}),\label{eq:att2} 
\end{align}
where 
$u_n$ is a decoder state vector
at $n$-th step, and $c_{n-1}$ is the context vector at the previous time step.
Then, context vector $c_n$ for the current time step $n$ is generated as a weighted sum of
the encoder embeddings as follows.
\begin{align}
 c_n&=\sum_{t=1}^T \alpha_{n,t}h^{enc}_t.
 \label{eq:context}
\end{align}
Finally, 
the output distribution for $y_n$ 
is estimated given
the context vector $c_n$
and decoder state vector $u_n$
as follows: 
\begin{align}
Pr(y_n|y_{1:n-1},X)&\sim {\rm DecoderOut}(c_n,u_n) \nonumber\\
&\hspace{-5mm}={\rm Softmax}(W_{out}\cdot {\rm LSTM}(c_n+u_n)). \label{eq:dec}
\end{align}
Here,
we are assuming 
that $c_n$ and $u_n$ have the same dimensionality.
Variable $W_{out}$ is the affine matrix of the final layer.
Note that
${\rm DecoderOut}$ normally consists of
a single affine transform with a softmax output layer.
However, it was found in \cite{kanda2020sot} that inserting one LSTM just before the affine transform
effectively improves the SOT model,
so we follow that architecture.

\subsection{Serialized Output Training}
With the SOT framework, 
the references for multiple overlapped utterances are concatenated to form a single token sequence
by inserting 
 a special symbol $\langle sc\rangle$
 representing a speaker change. 
For example, for the three-speaker case, the reference label will be given as
$R=\{r^1_{1},..,r^1_{N^1}, \langle sc\rangle, r^2_{1},..,r^2_{N^2}, \langle sc\rangle, r^3_{1},..,r^3_{N^3}, \langle eos\rangle\}$, 
where $r^j_i$ represents $i$-th token of $j$-th utterance.
Note that $\langle eos\rangle$, a token for sequence end,  is used only at the end of the entire sequence. 

Because there are multiple permutations in the order of reference labels
to form $R$,
some trick is needed to calculate the loss 
for AED.
One simple yet effective approach in \cite{kanda2020sot} is
sorting the reference labels
by their start times, which is called ``first-in, first-out'' (FIFO) training.
This training scheme works with complexity of $O(S)$ 
with respect to the number of speakers and 
outperforms a scheme that exhaustively considers 
all possible permutations \cite{kanda2020sot}.
In this paper, we always use this FIFO training
scheme.

Unlike permutation invariant training \cite{yu2017permutation,yu2017recognizing}, in
which 
the number of the output branches that a model has constrains
the number of recognizable speakers,
 SOT has no such limitation.
Refer to \cite{kanda2020sot} for a detailed description of the SOT model.

\section{Proposed method}
\subsection{Overview}

Suppose that we have a 
speaker inventory $\mathcal{D}=\{d_1,...,d_K\}$,
where $K$ is the number of speakers in the inventory
and $d_k$ is a speaker profile vector 
(e.g., d-vector \cite{variani2014deep}) 
of the $k$-th speaker.
The goal of the proposed method
 is 
 to estimate a
serialized multi-speaker transcription $Y$ 
accompanied by
the speaker identity of each token $S=\{s_1,...,s_N\}$ 
given
input 
$X$ and $\mathcal{D}$.

In this work, we 
 assume that
 the profiles of all
speakers involved in the input speech 
are included in $\mathcal{D}$. In other words,
we assume there is no ``unknown'' speaker 
for speaker identification.
As long as this condition holds,
the speaker inventory may include 
 any number of irrelevant speakers' profiles.
This is a typical setup in scheduled office meetings, where 
meeting organizers invite attendees whose voice profiles are pre-registered.

\begin{figure}[t]
  \centering
  \includegraphics[width=\linewidth]{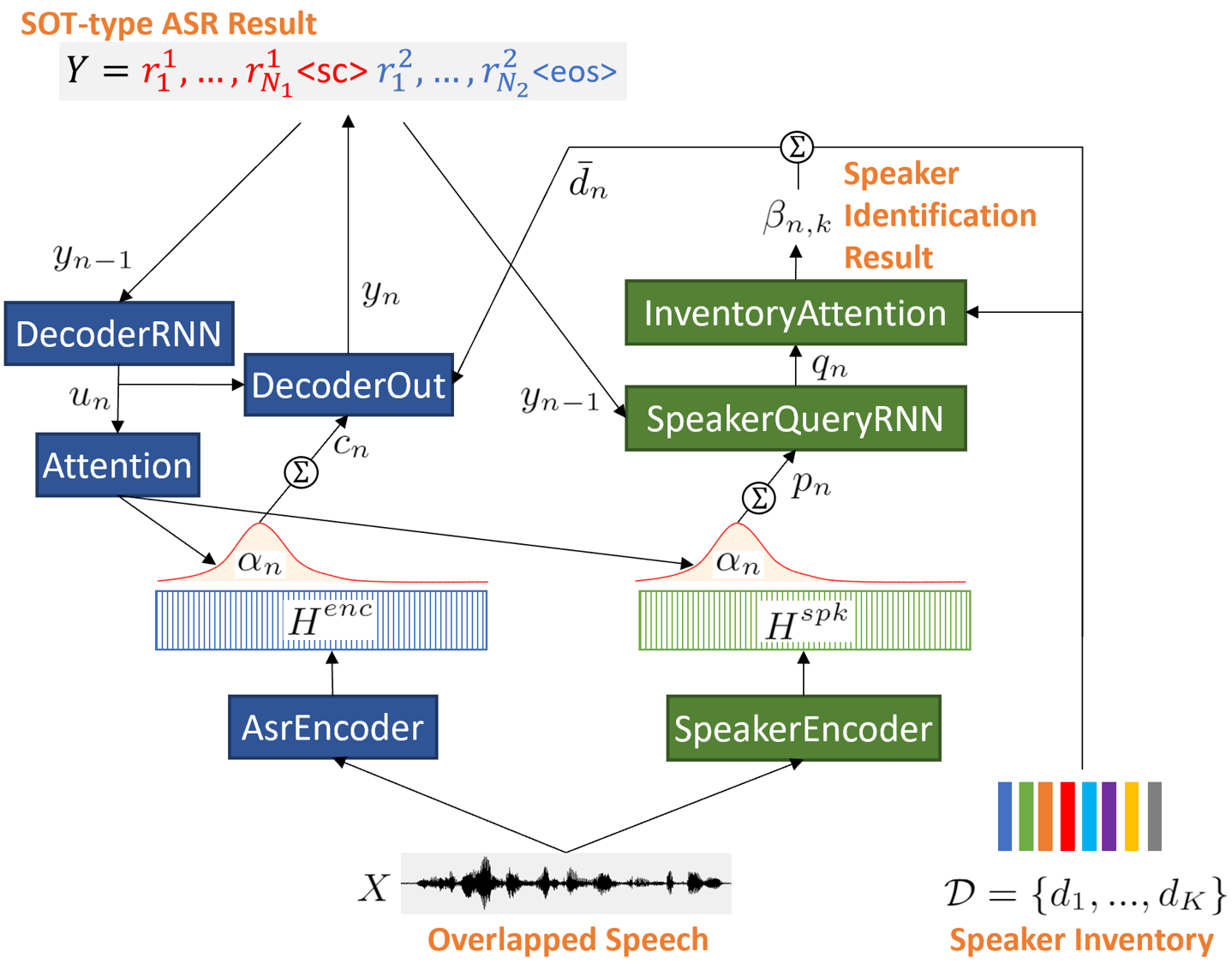}
  \vspace{-7mm}
  \caption{Proposed model.}
  \label{fig:asr_models}
  \vspace{-5mm}
\end{figure}

\subsection{Model Architecture}
We start with
 the conventional AED 
 represented by the blue blocks in Fig. \ref{fig:asr_models}.
Firstly, we introduce
one more encoder to represent the speaker characteristics of
input $X$ as follows.
 \begin{align}
 H^{spk} &= \{h^{spk}_1,...,h^{spk}_T\} = {\rm SpeakerEncoder}(X),  \label{eq:spkenc}
\end{align}
On top of this,
for every decoder step $n$,
we apply the attention weight $\alpha_n$ 
generated by the attention module
of AED
to extract attention-weighted vector of speaker embeddings
$p_n$.
\begin{align}
p_n
    &=\sum_{t=1}^T \alpha_{n,t}h^{spk}_t.
\end{align}
Note that 
$p_n$
could be contaminated by interfering speech
because some time frames include two or more speakers.

The speaker query RNN in Fig.~\ref{fig:asr_models} then generates a
speaker query $q_n$ given the speaker embedding 
$p_n$,
previous output $y_{n-1}$, and previous 
speaker query $q_{n-1}$.
\begin{align}
q_n &= {\rm SpeakerQueryRNN}(p_n,y_{n-1},q_{n-1}).
\end{align}
Based on the speaker query $q_n$,
an attention module for speaker inventory (shown as InventoryAttention in the diagram) 
estimates
the attention weight $\beta_{n,k}$ for each profile in $\mathcal{D}$.
\begin{align}
b_{n,k}&=\frac{q_n\cdot d_k}{|q_n||d_k|}, \label{eq:sp_att_cos}\\
\beta_{n,k}&= \frac{\exp(b_{n,k})}{\sum_j^K \exp(b_{n,j})}. 
\label{eq:sp_att_softmax}
\end{align}
Here, we use the softmax function (Eq. \eqref{eq:sp_att_softmax}) 
on the cosine similarity between the speaker query
 and speaker profile (Eq. \eqref{eq:sp_att_cos}), 
 which was found 
to be the most efficient in our preliminary experiment.
The attention weight $\beta_{n,k}$ can be seen as 
a posterior probability of speaker $k$ speaking the $n$-th token
given all previous tokens and speakers as well as $X$ and $\mathcal{D}$.
\begin{align}
Pr(s_n=k|y_{1:n-1},s_{1:n-1},X,\mathcal{D}) \sim\beta_{n,k}. \label{eq:spk-prob}
\end{align}

Finally, we calculate the attention-weighted 
speaker 
profile $\bar{d}_n$ 
based on $\beta_{n,k}$ and $d_k$ as followings.
\begin{align}
\bar{d}_n=\sum_{k=1}^{K}\beta_{n,k}d_k.
\end{align}
This weighted profile $\bar{d}_n$ is
appended to the input of $\rm DecoderOut()$. 
Specifically, we replace Eq. \eqref{eq:dec} by 
\begin{align}
Pr(y_n|y_{1:n-1},s_{1:n},X,\mathcal{D}) &\sim {\rm DecoderOut}(c_n,u_n,\bar{d}_n) \nonumber \\
&\hspace{-30mm}={\rm Softmax}(W_{out}\cdot {\rm LSTM}(c_n+u_n+W_d \bar{d}_n)), \label{eq:token-prob} 
\end{align}
where $W_d$ is a matrix to change the dimension of $\bar{d}_n$ to
that of $c_n$.
Terms $u_n$ and $c_n$ are obtained from Eq. \eqref{eq:att2} and \eqref{eq:context}, respectively. 
Note that the output distribution is now
conditioned on 
$s_{1:n}$ and
$\mathcal{D}$ because of the addition of 
the weighted profile 
$\bar{d}_n$.

\subsection{Training}
During training, all the network parameters are
optimized by maximizing
$\log Pr(Y,S|X,\mathcal{D})$ as follows.
We call it {\it speaker-attributed maximum mutual information (SA-MMI)} training.
\begin{align}
\mathcal{F}^{\mathrm{SA-MMI}}&=\log Pr(Y,S|X,\mathcal{D}) \label{eq:samll-1}\\
&\hspace{-10mm}= \log \prod_{n=1}^{N}\{Pr(y_{n}|y_{1:n-1}, s_{1:n}, X, \mathcal{D}) \nonumber \\ 
&\hspace{-10mm} \;\;\;\;\;\;\;\;\;\;\;\;\;\;\; \cdot Pr(s_{n}|y_{1:n-1}, s_{1:n-1}, X, \mathcal{D})^\gamma \} \label{eq:samll-2}\\
&\hspace{-10mm}=\sum_{n=1}^{N}\log Pr(y_{n}|y_{1:n-1}, s_{1:n}, X, \mathcal{D}) \nonumber \\ 
&\hspace{-10mm} \;\;\;\;  +\gamma\cdot\sum_{n=1}^{N} \log Pr(s_{n}|y_{1:n-1}, s_{1:n-1}, X, \mathcal{D}). \label{eq:samll-3} 
\end{align}
From Eq. \eqref{eq:samll-1} to Eq. \eqref{eq:samll-2}, 
the chain rule is applied for
$y_n$ and $s_n$ alternately. 
Here, we 
introduce a scaling parameter $\gamma$ to adjust the scale of the speaker estimation probability to that of ASR. 
Equation \eqref{eq:samll-3} shows that our 
training criterion can be factorized into
two conditional probabilities defined in
Eqs. \eqref{eq:token-prob} and \eqref{eq:spk-prob}, 
respectively.
Note that 
 the speaker identity of the token
 $\langle sc\rangle$ or
 $\langle eos\rangle$  
 is
set the
same as
that 
of 
the preceding token.

\subsection{Decoding}
An extended beam search algorithm is used for decoding with 
the proposed method.
In the conventional beam search for AED,
each hypothesis 
contains estimated tokens accompanied by the posterior probability of 
the hypothesis.
In addition to these,
a hypothesis for the proposed method contains
speaker estimation $\beta_{n,k}$.
 Each hypothesis expands until $\langle eos\rangle$ is detected,
 and the estimated tokens in each hypothesis are grouped by  
 $\langle sc\rangle$ 
 to form multiple utterances.
 For each utterance,
  the average of  $\beta_{n,k}$ values, including the last token corresponding to 
   $\langle sc\rangle$
   or $\langle eos\rangle$, 
is calculated for each speaker. 
The speaker with the highest average $\beta_{n,k}$ score is 
 selected as the predicted speaker of that utterance.
 Finally, when the same speaker is predicted for multiple
 utterances, those utterances are concatenated to form a single utterance.

\begin{table*}[t]
  \caption{SER (\%), WER (\%), and SA-WER (\%) for  
baseline systems and proposed method. 
The number of profiles per test audio was 8. 
Each profile was extracted by using  2 utterances (15 sec on average). 
For random speaker assignment experiment (3rd row),  averages of 10 trials were computed. 
No LM was used in the evaluation.}
  \label{tab:baseline}
  \vspace{-3mm}
  \centering
{\scriptsize
  \begin{tabular}{c|ccc|ccc|ccc|ccc}
    \toprule
\multirow{2}{*}{\backslashbox{Model}{Eval Set}} &  \multicolumn{3}{c|}{1-speaker} & \multicolumn{3}{c|}{2-speaker-mixed } & \multicolumn{3}{c|}{3-speaker-mixed} & \multicolumn{3}{c}{Total}\\
 &  SER & WER & {\bf SA-WER} &  SER & WER & {\bf SA-WER} &  SER & WER & {\bf SA-WER} &   SER &  WER & {\bf SA-WER}  \\
    \midrule
\multicolumn{1}{l|}{Single-speaker ASR} & - & 4.7 & - & - & 66.9 & - & - & 90.7 & - & - &  68.4 & - \\ 
\multicolumn{1}{l|}{SOT-ASR} & - & 4.5 & - & - & 10.3 & - & - & 19.5 & - & - &  13.9 & - \\ \midrule
\multicolumn{1}{l|}{SOT-ASR + random speaker assignment} & 87.4 & 4.5 & {\bf 175.2} & 82.8 &  23.4 & {\bf 169.7} & 76.1 & 39.1 & {\bf 165.1} & 80.2 & 28.1 & {\bf 168.3} \\ 
\multicolumn{1}{l|}{SOT-ASR + d-vec speaker identification} & 0.4 & 4.5 & {\bf 4.8} & 6.4 & 10.3 & {\bf 16.5}  & 13.1 & 19.5 & {\bf 31.7} & 8.7  & 13.9  & {\bf 22.2} \\ \midrule
\multicolumn{1}{l|}{\underline{Proposed Model}} & &  &   &  &  &  &  &  &  & & &  \\ 
\multicolumn{1}{l|}{SOT-ASR + Spk-Enc + Inv-Attn} & 0.3 & 4.3 & {\bf 4.7}  & 5.5 & 10.4 & {\bf 12.2} & 14.8 & 23.4 & {\bf 26.7} &  9.3 & 15.9 & {\bf 18.2} \\ 
\multicolumn{1}{r|}{$\hookrightarrow$ + SpeakerQueryRNN $\;\;\;\;\;\;\;\;$} & 0.4 & 4.2 & {\bf 4.6} & 3.0 & 9.1 & {\bf 10.9} & 11.6 & 21.5 & {\bf 24.7} &  6.9 &   14.5 & {\bf 16.7} \\ 
\multicolumn{1}{r|}{$\hookrightarrow$ + Weighted Profile ($\bar{d}_n$)} & 0.2 & 4.2 & {\bf 4.5}  & 2.5 & 8.7 & {\bf 9.9} & 10.2 & 20.2 & {\bf 23.1} &  6.0 &  13.7 & {\bf 15.6} \\ 
     \bottomrule
  \end{tabular}
}
  \vspace{-5mm}
\end{table*}

\begin{table}[t]
  \caption{Speaker counting accuracy (\%) of the proposed model.}
  \label{tab:spk-count}
  \vspace{-3mm}
  \centering
  {\scriptsize
  \begin{tabular}{c|cccc}
    \toprule
Actual \# of Speakers& \multicolumn{4}{c}{Estimated \# of Speakers (\%)} \\
in Test Data & 1 & 2 & 3 & $>$4 \\
    \midrule
 1 & {\bf 99.96} & 0.04 & 0.00 & 0.00\\ \midrule
 2 & 2.56 & {\bf 97.44} & 0.00 & 0.00 \\ \midrule
 3 & 0.31 & 25.34 & {\bf 74.35} & 0.00\\ 
    \bottomrule
  \end{tabular}
  }
  \vspace{-3mm}
\end{table}

\begin{table}[t]
  \caption{SER (\%) / SA-WER (\%) for different numbers of profiles in the speaker inventory. }
  \label{tab:profilesize}
  \vspace{-3mm}
  \centering
  {\scriptsize
  \begin{tabular}{c|ccc|c}
    \toprule
\# of Profiles &   \multicolumn{3}{c}{\# of Speakers in Test Data}& \\
in Inventory  & 1 & 2 & 3 & {\bf Total}\\
    \midrule
  4  & 0.1 / 4.5 & 1.8 / 9.4  & 8.8 / 22.3 & {\bf 5.0 / 15.0} \\ 
  8  & 0.2 / 4.5 & 2.5 / 9.9  & 10.2 / 23.1  & {\bf 6.0 / 15.6} \\ 
  16 & 0.8 / 5.1 & 2.9 / 10.6 & 11.3 / 23.8  & {\bf 6.8 / 16.3} \\ 
  32 & 0.9 / 5.4 & 4.6 / 11.9 & 11.6 / 24.0  & {\bf 7.5 / 16.9} \\ 
    \bottomrule
  \end{tabular}
  }
  \vspace{-5mm}
\end{table}

\begin{table}[t]
  \caption{Impact of the number of profile extraction utterances on SER (\%) and SA-WER (\%) for 8-profile setting. 
The average utterance duration 
    was 7.5 sec.} 
  \label{tab:profilequality}
  \vspace{-3mm}
  \centering
  {\scriptsize
  \begin{tabular}{c|ccc|c}
    \toprule
\# of Utterances &    \multicolumn{3}{c}{\# of Speakers in Test Data} & \\
per Profile &  1 & 2 & 3 & {\bf Total}\\
    \midrule
 1  & 0.9 / 5.6 & 3.8 / 11.5 & 11.2 / 24.8 & {\bf 7.0 / 17.2} \\ 
 2  & 0.2 / 4.5 & 2.5 / 9.9 & 10.2 / 23.1 & {\bf 6.0 / 15.6} \\ 
 5  & 0.04 / 4.2 & 2.1 / 9.5 & 9.7 / 22.6 & {\bf 5.6 / 15.2} \\ 
 10 & 0.08 / 4.3 & 2.0 / 9.4 & 9.5 / 22.3 & {\bf 5.4 / 15.0} \\ 
    \bottomrule
  \end{tabular}
  }
  \vspace{-5mm}
\end{table}

\section{Experiments}
\subsection{Evaluation settings}
\subsubsection{Evaluation data}
We evaluated the effectiveness of the proposed method by 
simulating multi-speaker signals based on the LibriSpeech corpus~\cite{panayotov2015librispeech}.
Following the Kaldi \cite{povey2011kaldi} recipe, 
we used
the 960 hours of LibriSpeech training data (``train\_960'') for model learning,
the ``dev\_clean'' set for adjusting hyper-parameter values, 
and the ``test\_clean'' set for testing. 

Our training data were generated as follows. 
For each utterance in train\_960, 
randomly chosen $(S-1)$ train\_960 utterances were added after being shifted by random delays. 
When mixing the audio signals, the original volume of each utterance was kept unchanged, resulting in an average signal-to-interference ratio of about 0 dB. 
As for the delay applied to each utterance, the delay values were randomly chosen under the constraints 
that (1) the start times of the individual utterances differed by 0.5 sec or longer
and that (2) every utterance in each mixed audio sample had at least one speaker-overlapped region with other utterances.
For each training sample, speaker profiles were generated as follows. 
First, the number of profiles was randomly selected 
from  $S$ 
to  8. 
Among those profiles, $S$ profiles were for the speakers involved in the overlapped speech. 
The utterances for creating the profiles of these speakers were different from those constituting the input overlapped speech. 
The rest of the profiles were randomly extracted from the other speakers in train\_960. 
Each profile was extracted by using 
10 utterances. 
We generated data for $S=\{1,2,3\}$ and 
combined them to use for training.

The development and evaluation sets were generated from dev\_clean or test\_clean, respectively, in the 
same way as the training set except that 
constraint (1) was not imposed. 
Therefore, 
multiple utterances were allowed to start at the same time
in evaluation.
Also, each profile was extracted from 
2 utterances (15 sec on average) instead of 10, 
unless otherwise stated. 

\subsubsection{Evaluation metric}
We evaluated the model with respect to 
 speaker error rate (SER),
WER, and SA-WER. 
{\bf SER} is defined as the total number of 
model-generated utterances with speaker misattribution 
divided by
the number of reference utterances. 
All possible permutations of the hypothesized utterances were examined
by ignoring the ASR results,
and the one that yielded the smallest number of errors
(including the speaker insertion and deletion errors) was picked for the SER calculation. 
Similarly, {\bf WER} was calculated 
by picking the best permutation in terms of word errors (i.e., speaker labels were ignored). 
Finally, {\bf SA-WER} was calculated 
by comparing the ASR hypothesis and the reference transcription of each speaker.

\subsubsection{Model settings}
In our experiments,
we used a 80-dim log mel filterbank, extracted every 10 msec, for the input feature.
We stacked 3 frames of features and applied the model
on top of the stacked features.
For the speaker profile, we used
a 128-dim d-vector \cite{variani2014deep}, 
whose extractor was separately 
trained on VoxCeleb Corpus \cite{nagrani2017voxceleb,chung2018voxceleb2}.
The d-vector extractor 
consisted of 17 convolution layers followed by an average pooling layer, which was a modified version of the one presented in \cite{zhou2019cnn}. 

Our AsrEncoder consisted of
 5 layers of 1024-dim 
bidirectional long short-term memory (BLSTM), interleaved with layer normalization~\cite{ba2016layer}.
The DecoderRNN consisted of 
2 layers of 1024-dim unidirectional LSTM,
and the DecoderOut consisted of 1 layer of 1024-dim unidirectional LSTM. 
We used a conventional location-aware content-based attention \cite{chorowski2015attention} with 
a single attention head. 
The SpeakerEncoder had the same architecture as the d-vector extractor except for not having the final average pooling layer.
Our SpeakerQueryRNN consisted of 1 layer of 512-dim unidirectional LSTM.
We used 16k subwords based on a unigram language model \cite{kudo2018subword}
as a recognition unit. 
We appplied volume perturbation to the mixed audio to
increase the training data variability.
Note that
we applied neither 
an additional language model (LM) nor
any other forms of 
data augmentation 
\cite{kanda2013elastic,ko2015audio,park2019specaugment,wang2019semantic}
for simplicity. 

Model training was performed as follows. 
In our preliminary experiment, 
training models from fully random parameters 
showed poor convergence due to the difficulty in 
attention module training. 
Therefore, we initialized the parameters of
AsrEncoder, Attention, DecoderRNN, and DecoderOut
by SOT-ASR parameters trained on  
simulated mixtures of LibriSpeech utterances as reported in \cite{kanda2020sot}.
We pre-trained the SOT-model with 640k iterations.
We also initialized the SpeakerEncoder parameters 
by using those of the d-vector extractor. 
After the initialization,
we updated the entire network based on $\mathcal{F}^{\mathrm{SA-MMI}}$
with $\gamma=0.1$
by using
an Adam optimizer
with a learning rate of 0.00002.
We used 8 GPUs, each of which worked
on 6k frames of minibatch.
We report the results 
of the dev\_clean-based best models found after 160k of training iterations.

\subsection{Evaluation results}

\subsubsection{Baseline results}

We built 4 different baseline systems, whose results are shown 
in the first 4 rows of Table \ref{tab:baseline}.
The first row corresponds the conventional 
single-speaker ASR based on AED.
As expected, the WER was significantly degraded 
for overlapped speech.
The second row shows the result of the SOT-ASR system 
that was used for initializing the proposed method in training. 
SOT-ASR significantly improved the WER 
for all evaluation settings.
The lower WER for the 1-speaker case could be attributed to the data augmentation effect resulting from the use of 
overlapped speech for training, 
which was also observed in \cite{kanda2020sot}.

The third row shows the result of randomly assigning
a speaker label for each utterance generated by SOT-ASR. 
Note that the speaker identification may affect 
WER as well as SA-WER.
This is because 
multiple SOT-ASR-generated utternaces were mereged 
when their speaker labels were the same. 

The fourth row shows the result of 
combining SOT-ASR and d-vector based speaker identification.
In this baseline system,
for each utterance, 
we calculated a weighted average of frame-level d-vectors by using
the attention weights from SOT-ASR.
The estimated d-vectors were then compared with each profile
contained in the speaker inventory in terms of cosine similarity.
The best scored speaker was selected one-by-one 
with a constraint
that the same speaker could  not be selected for multiple utterances.
This method gave us reasonable results as can be seen in the table 
 although the SA-WERs were not sufficient for overlapped speech.

\subsubsection{Results of the proposed method}
The last 3 rows of Table \ref{tab:baseline} shows
the results of the proposed method 
while the first two of them 
were the results of an ablation study. 
``SOT-ASR + Spk-Enc + Inv-Attn'' is
the result of a variant 
of the proposed model where 
$\rm SpeakerEncoder$ output $p_n$ was directly used for $\rm InventoryAttention$ (Eq. \eqref{eq:sp_att_cos}) 
instead of 
using $\rm SpeakerQueryRNN$ output
$q_n$, 
and $\bar{d}_n$ was not used
in Eq. \eqref{eq:token-prob}.
Due to SA-MMI training,
even this model achieved a  lower SA-WER
than the baseline 
while the SER and WER were 
degraded.
Then, the entire performance was
significantly boosted by 
introducing
$\rm SpeakerQueryRNN$
as shown in the next row.
Finally, 
by introducing the weighted profile $\bar{d}_n$ in Eq. \eqref{eq:token-prob}, 
the proposed method outperformed the baseline
in all three evaluation metrics,
resulting in
29\% reduction of the SA-WER.

Table \ref{tab:spk-count} shows the speaker counting accuracy of the proposed method.
We can see that the speakers were counted very accurately  especially for the 1-speaker  (99.96\%) 
and  2-speaker  cases  (97.44\%) 
while  it  sometimes 
underestimated the number of the speakers for the 3-speaker mixtures.

\subsubsection{Evaluation with different profile settings}

In the previous experiments, we used the inventory comprising 
8 profiles,
each of which was
extracted from 2 utterances. 
We then evaluated the proposed method with different
numbers of profiles.
As shown in Table \ref{tab:profilesize}, our proposed method
showed only minor degradation in terms of the SER and SA-WER
even with 32 profiles. 
This demonstrates the 
robustness of our method against the increase of profiles.

Finally, we also evaluated 
the impact of the number of utterances used 
for speaker profile extraction.
As shown in Table \ref{tab:profilequality},
using more utterances for a profile yielded 
lower error rates.

\section{Conclusions}

In this paper, we proposed
a joint model
for SA-ASR
that can recognize
overlapped speech of any number of speakers 
while identifying the speaker of
each utterance among any number of speaker profiles.
In the experiments on LibriSpeech, 
 the proposed model achieved  
 significantly  better  SA-WER
than the baseline that consists of separated modules.

\bibliographystyle{IEEEtran}

\bibliography{mybib}

\begin{thebibliography}{10}
\providecommand{\url}[1]{#1}
\csname url@samestyle\endcsname
\providecommand{\newblock}{\relax}
\providecommand{\bibinfo}[2]{#2}
\providecommand{\BIBentrySTDinterwordspacing}{\spaceskip=0pt\relax}
\providecommand{\BIBentryALTinterwordstretchfactor}{4}
\providecommand{\BIBentryALTinterwordspacing}{\spaceskip=\fontdimen2\font plus
\BIBentryALTinterwordstretchfactor\fontdimen3\font minus
  \fontdimen4\font\relax}
\providecommand{\BIBforeignlanguage}[2]{{%
\expandafter\ifx\csname l@#1\endcsname\relax
\typeout{** WARNING: IEEEtran.bst: No hyphenation pattern has been}%
\typeout{** loaded for the language `#1'. Using the pattern for}%
\typeout{** the default language instead.}%
\else
\language=\csname l@#1\endcsname
\fi
#2}}
\providecommand{\BIBdecl}{\relax}
\BIBdecl

\bibitem{fiscus2007rich}
J.~G. Fiscus, J.~Ajot, and J.~S. Garofolo, ``The rich transcription 2007
  meeting recognition evaluation,'' in \emph{Multimodal Technologies for
  Perception of Humans}.\hskip 1em plus 0.5em minus 0.4em\relax Springer, 2007,
  pp. 373--389.

\bibitem{janin2003icsi}
A.~Janin, D.~Baron, J.~Edwards, D.~Ellis, D.~Gelbart, N.~Morgan, B.~Peskin,
  T.~Pfau, E.~Shriberg, A.~Stolcke \emph{et~al.}, ``The {ICSI} meeting
  corpus,'' in \emph{Proc. ICASSP}, vol.~1, 2003, pp. I--I.

\bibitem{carletta2005ami}
J.~Carletta, S.~Ashby, S.~Bourban, M.~Flynn, M.~Guillemot, T.~Hain, J.~Kadlec,
  V.~Karaiskos, W.~Kraaij, M.~Kronenthal \emph{et~al.}, ``The {AMI} meeting
  corpus: A pre-announcement,'' in \emph{International workshop on machine
  learning for multimodal interaction}.\hskip 1em plus 0.5em minus 0.4em\relax
  Springer, 2005, pp. 28--39.

\bibitem{yoshioka2019advances}
T.~Yoshioka, I.~Abramovski, C.~Aksoylar, Z.~Chen, M.~David, D.~Dimitriadis,
  Y.~Gong, I.~Gurvich, X.~Huang, Y.~Huang \emph{et~al.}, ``Advances in online
  audio-visual meeting transcription,'' in \emph{Proc. ASRU}, 2019, pp.
  276--283.

\bibitem{hershey2016deep}
J.~R. Hershey, Z.~Chen, J.~Le~Roux, and S.~Watanabe, ``Deep clustering:
  Discriminative embeddings for segmentation and separation,'' in \emph{Proc.
  ICASSP}, 2016, pp. 31--35.

\bibitem{chen2017deep}
Z.~Chen, Y.~Luo, and N.~Mesgarani, ``Deep attractor network for
  single-microphone speaker separation,'' in \emph{Proc. ICASSP}, 2017, pp.
  246--250.

\bibitem{yu2017permutation}
D.~Yu, M.~Kolb{\ae}k, Z.-H. Tan, and J.~Jensen, ``Permutation invariant
  training of deep models for speaker-independent multi-talker speech
  separation,'' in \emph{Proc. ICASSP}.\hskip 1em plus 0.5em minus 0.4em\relax
  IEEE, 2017, pp. 241--245.

\bibitem{yu2017recognizing}
D.~Yu, X.~Chang, and Y.~Qian, ``Recognizing multi-talker speech with
  permutation invariant training,'' \emph{Proc. Interspeech 2017}, pp.
  2456--2460, 2017.

\bibitem{seki2018purely}
H.~Seki, T.~Hori, S.~Watanabe, J.~Le~Roux, and J.~R. Hershey, ``A purely
  end-to-end system for multi-speaker speech recognition,'' in \emph{Proc.
  ACL}, 2018, pp. 2620--2630.

\bibitem{chang2019end}
X.~Chang, Y.~Qian, K.~Yu, and S.~Watanabe, ``End-to-end monaural multi-speaker
  {ASR} system without pretraining,'' in \emph{Proc. ICASSP}, 2019, pp.
  6256--6260.

\bibitem{chang2019mimo}
X.~Chang, W.~Zhang, Y.~Qian, J.~L. Roux, and S.~Watanabe, ``{MIMO-SPEECH}:
  End-to-end multi-channel multi-speaker speech recognition,'' in \emph{Proc.
  ASRU}, 2019, pp. 237--244.

\bibitem{kanda2019acoustic}
N.~Kanda, Y.~Fujita, S.~Horiguchi, R.~Ikeshita, K.~Nagamatsu, and S.~Watanabe,
  ``Acoustic modeling for distant multi-talker speech recognition with
  single-and multi-channel branches,'' in \emph{Proc. ICASSP}, 2019, pp.
  6630--6634.

\bibitem{kanda2019auxiliary}
N.~Kanda, S.~Horiguchi, R.~Takashima, Y.~Fujita, K.~Nagamatsu, and S.~Watanabe,
  ``Auxiliary interference speaker loss for target-speaker speech
  recognition,'' in \emph{Proc. Interspeech}, 2019, pp. 236--240.

\bibitem{wang2019speech}
P.~Wang, Z.~Chen, X.~Xiao, Z.~Meng, T.~Yoshioka, T.~Zhou, L.~Lu, and J.~Li,
  ``Speech separation using speaker inventory,'' in \emph{Proc. ASRU}, 2019,
  pp. 230--236.

\bibitem{von2019all}
T.~von Neumann, K.~Kinoshita, M.~Delcroix, S.~Araki, T.~Nakatani, and
  R.~Haeb-Umbach, ``All-neural online source separation, counting, and
  diarization for meeting analysis,'' in \emph{Proc. ICASSP}, 2019, pp. 91--95.

\bibitem{kinoshita2020tackling}
K.~Kinoshita, M.~Delcroix, S.~Araki, and T.~Nakatani, ``Tackling real noisy
  reverberant meetings with all-neural source separation, counting, and
  diarization system,'' \emph{arXiv preprint arXiv:2003.03987}, 2020.

\bibitem{el2019joint}
L.~El~Shafey, H.~Soltau, and I.~Shafran, ``Joint speech recognition and speaker
  diarization via sequence transduction,'' in \emph{Proc. Interspeech}, 2019,
  pp. 396--400.

\bibitem{kanda2019simultaneous}
N.~Kanda, S.~Horiguchi, Y.~Fujita, Y.~Xue, K.~Nagamatsu, and S.~Watanabe,
  ``Simultaneous speech recognition and speaker diarization for monaural
  dialogue recordings with target-speaker acoustic models,'' in \emph{Proc.
  ASRU}, 2019.

\bibitem{kanda2020sot}
N.~Kanda, Y.~Gaur, X.~Wang, Z.~Meng, and T.~Yoshioka, ``Serialized output
  training for end-to-end overlapped speech recognition,'' \emph{arXiv preprint
  arXiv:2003.12687}, 2020.

\bibitem{bahdanau2014neural}
D.~Bahdanau, K.~Cho, and Y.~Bengio, ``Neural machine translation by jointly
  learning to align and translate,'' \emph{arXiv preprint arXiv:1409.0473},
  2014.

\bibitem{chorowski2014end}
J.~Chorowski, D.~Bahdanau, K.~Cho, and Y.~Bengio, ``End-to-end continuous
  speech recognition using attention-based recurrent {NN}: First results,'' in
  \emph{NIPS Workshop on Deep Learning}, 2014.

\bibitem{chorowski2015attention}
J.~K. Chorowski, D.~Bahdanau, D.~Serdyuk, K.~Cho, and Y.~Bengio,
  ``Attention-based models for speech recognition,'' in \emph{Proc. NIPS},
  2015, pp. 577--585.

\bibitem{chan2016listen}
W.~Chan, N.~Jaitly, Q.~Le, and O.~Vinyals, ``Listen, attend and spell: A neural
  network for large vocabulary conversational speech recognition,'' in
  \emph{Proc. ICASSP}, 2016, pp. 4960--4964.

\bibitem{variani2014deep}
E.~Variani, X.~Lei, E.~McDermott, I.~L. Moreno, and J.~Gonzalez-Dominguez,
  ``Deep neural networks for small footprint text-dependent speaker
  verification,'' in \emph{Proc. ICASSP}, 2014, pp. 4052--4056.

\bibitem{panayotov2015librispeech}
V.~Panayotov, G.~Chen, D.~Povey, and S.~Khudanpur, ``Librispeech: an {ASR}
  corpus based on public domain audio books,'' in \emph{Proc. ICASSP}, 2015,
  pp. 5206--5210.

\bibitem{povey2011kaldi}
D.~Povey, A.~Ghoshal, G.~Boulianne, L.~Burget, O.~Glembek, N.~Goel,
  M.~Hannemann, P.~Motlicek, Y.~Qian, P.~Schwarz \emph{et~al.}, ``The {Kaldi}
  speech recognition toolkit,'' in \emph{ASRU}, 2011.

\bibitem{nagrani2017voxceleb}
A.~Nagrani, J.~S. Chung, and A.~Zisserman, ``Voxceleb: A large-scale speaker
  identification dataset,'' in \emph{Proc. Interspeech}, 2017, pp. 2616--2620.

\bibitem{chung2018voxceleb2}
J.~S. Chung, A.~Nagrani, and A.~Zisserman, ``Voxceleb2: Deep speaker
  recognition,'' in \emph{Proc. Interspeech}, 2018, pp. 1086--1090.

\bibitem{zhou2019cnn}
T.~Zhou, Y.~Zhao, J.~Li, Y.~Gong, and J.~Wu, ``{CNN} with phonetic attention
  for text-independent speaker verification,'' in \emph{Proc. ASRU}, 2019, pp.
  718--725.

\bibitem{ba2016layer}
J.~L. Ba, J.~R. Kiros, and G.~E. Hinton, ``Layer normalization,'' \emph{arXiv
  preprint arXiv:1607.06450}, 2016.

\bibitem{kudo2018subword}
T.~Kudo, ``Subword regularization: Improving neural network translation models
  with multiple subword candidates,'' \emph{arXiv preprint arXiv:1804.10959},
  2018.

\bibitem{kanda2013elastic}
N.~Kanda, R.~Takeda, and Y.~Obuchi, ``Elastic spectral distortion for low
  resource speech recognition with deep neural networks,'' in \emph{Proc.
  ASRU}, 2013, pp. 309--314.

\bibitem{ko2015audio}
T.~Ko, V.~Peddinti, D.~Povey, and S.~Khudanpur, ``Audio augmentation for speech
  recognition,'' in \emph{Proc. Interspeech}, 2015, pp. 3586--3589.

\bibitem{park2019specaugment}
D.~S. Park, W.~Chan, Y.~Zhang, C.-C. Chiu, B.~Zoph, E.~D. Cubuk, and Q.~V. Le,
  ``Specaugment: A simple data augmentation method for automatic speech
  recognition,'' in \emph{Proc. Interspeech}, 2019, pp. 2613--2617.

\bibitem{wang2019semantic}
C.~Wang, Y.~Wu, Y.~Du, J.~Li, S.~Liu, L.~Lu, S.~Ren, G.~Ye, S.~Zhao, and
  M.~Zhou, ``Semantic mask for transformer based end-to-end speech
  recognition,'' \emph{arXiv preprint arXiv:1912.03010}, 2019.

\end{thebibliography}

\end{document}